\documentclass[nofootinbib,showpacs,preprintnumbers,amsmath,amssymb,floatfix]
{revtex4-1}
\usepackage{graphicx}
\usepackage{wrapfig}

\begin{document}

\title{Elastic scattering of virtual photons via quark loop in Double-Logarithmic Approximation}

\vspace*{0.3 cm}

\author{B.I.~Ermolaev}
\affiliation{Ioffe Physico-Technical Institute, 194021}
\author{D.Yu.~Ivanov}
\affiliation{Sobolev Institute of Mathematics, 630090 Novosibirsk, Russia}
 \author{S.I.~Troyan}
\affiliation{St.Petersburg Institute of Nuclear Physics, 188300
Gatchina, Russia}

\begin{abstract}
We calculate the amplitude of elastic photon-photon scattering via a single quark loop in the
Double-Logarithmic Approximation, presuming all external photons to be off-shell and unpolarized..
At the same time we account for the running coupling effects.
We consider this process in the forward kinematics at arbitrary relations between $t$ and the
external photon virtualities.
 We obtain explicit expressions for the photon-photon scattering
amplitudes in all double-logarithmic kinematic regions. Then we calculate the small-$x$ asymptotics
of the obtained amplitudes and compare them with the parent amplitudes, thereby
 fixing the  applicability regions of the asymptorics, i.e. fixing the applicability region for the non-vacuum Reggeons.
 We find that these Reggeons should be used at $x < 10^{-8}$ only.
\end{abstract}


\maketitle

\section{Introduction}

Since long the light-by-light scattering has been an object of both experimental and theoretical interest.
In this paper we consider this process in the high energy limit. The motivation of our study is twofold.

On the one hand, it is well known that, similarly to the $e^+e^-$
annihilation into hadrons, the total cross section of collision of
two off-shell photons with large virtualities is an important test ground
for perturbative QCD. At a fixed order of in the strong coupling, $\alpha_s$,
and at low energies, the dominant contribution comes from the pure QED
quark box diagrams, calculated at the leading-order (LO) in Refs. \cite{Budnev:1974de,Schienbein:2002wj}
and at the next-to-LO (NLO) in $\alpha_s$, see Ref. \cite{Cacciari:2000cb}.
In Ref. \cite{BL03} the resummation of double logs appearing starting from the first
NLO QCD corrections to the quark box was studied. Such contribution are important at high energy where arguments of the logs are large. At even higher energies additional class of QCD diagrams gives important contribution to the cross section. It is a contribution with the two-gluon exchange
in the t-channel that overwhelms the quark exchange contribution despite additional $\alpha_s^2$ suppression: it has a different asymptotics in the power of energy and therefore  it will exceed the contribution of   quark exchange mechanism at sufficiently  large  c.m.f. energy $\sqrt{s}$. At higher orders in $\alpha_s$, the contributions from
t-channel gluons lead
to terms with powers of single logarithms of the energy, which  must be resummed.
The BFKL approach \cite{BFKL} provides for a consistent theoretical framework for such resummation  of the energy logarithms, both in the leading logarithmic approximation (LLA),
which means resummation of all terms $\sim (\alpha_s\ln(s))^n$, and in the next-to-leading approximation
(NLA), which means resummation of all terms $\sim \alpha_s(\alpha_s\ln(s))^n$. In this approach, the imaginary
part of the amplitude (and, hence, the total cross section) for a large-$s$
hard collision process
can be written as the convolution of the Green’s function of two interacting Reggeized gluons with the impact factors of the colliding particles.

The study of the $\gamma^* \gamma^*$ total cross section in LLA BFKL has a long
history~\cite{photons_BFKL}. For the extension of these results to the NLA
level one needs to consider corrections to both the BFKL Green's function \cite{NLA-kernel} and
to the impact factors of colliding virtual photons.

While its LO expression for the photon impact factor is known since long, the NLO
calculation, carried out in the momentum representation, turned out to be
rather complicated and was completed only after year-long
efforts~\cite{gammaIF}, and the results are available only in the form of a numerical code,
thus making it of limited practical use. Indeed, until  recently,
the inclusion of BFKL resummation effects in the NLA calculation of the
$\gamma^* \gamma^*$ total cross section was carried out only in approximate
way, by taking the BFKL Green's function in the NLA while using the LO
expression for impact factors. This is the case of the pioneer paper in
Ref.~\cite{Brodsky:2002ka} (see also Ref.~\cite{Brodsky:1998kn}) and of the
later analysis in Refs.~\cite{Caporale2008} and~\cite{Zheng}.

The situation changed  when the NLO photon impact
was calculated analytically in the coordinate space and then transformed to
the momentum representation and to the Mellin ~\cite{Balitsky2012} (see also
Ref.~\cite{Chirilli2014}). This achievement opened a way for a subsequent calculations of
$\gamma^* \gamma^*$ total cross section with complete NLA BFKL resummation approach, see \cite{Chirilli2014, Ivanov:2014hpa}.

In \cite{Ivanov:2014hpa} a comparison of the NLA BFKL predictions with LEP2
data\cite{Achard:2001kr,Abbiendi:2001tv} was made. It was shown that the account of the Balitsky and Chirilli expression for
NLO photon impact factor reduces the BFKL contribution to the cross section
to very small values making it impossible to describe LEP2 data as a sum of
BFKL and LO QED quark box contributions. Note that, as we discussed above, the LO QED quark box
itself receives, at higher QCD orders, large corrections enhanced by double
logs. Their resummation is important and leads to a considerable enhancement
of the quark box contribution (see Ref.~\cite{BL03} for detail),
but still these effects are not large enough for a good description of LEP2
data at largest available rapidity without a sizable BFKL contribution.
Therefore, in this situation,  one of the aim of this paper is to reconsider the
derivation of double logs resummation and to confirm results of \cite{BL03}.
Besides, we account for
the running QCD coupling effects.

Another motivation for the present paper is related with the possibility to measure amplitude of
the light-by-light scattering at non zero angles, i.e. at non-zero values of $t$.
Recently,
 ATLAS Collaboration has  reported\cite{atlas} on evidence for the quasi-real photo-photon scattering scattering in heavy-ion collisions with the
ATLAS detector at the LHC.
These results proved to be consistent with calculations reported in Ref.~\cite{2,3,4},
Light-by-light scattering has been an object of both experimental and theoretical interest. For instance,
 ATLAS Collaboration has recently reported\cite{atlas} on evidence for light-by-light scattering in heavy-ion collisions with the
ATLAS detector at the LHC. These results proved to be consistent with calculations reported in Ref.~\cite{2,3,4},
where one of the essential ingredients is the amplitudes of the photon-photon elastic scattering studied in
the lowest ("Born") approximation where description of the photon scattering involves a single quark loop only.
As accounting for the QCD radiative corrections can essentially change the scattering amplitudes, it
is interesting to study their impact. Both technology of accounting for the radiative corrections and their impact
strongly depend on the kinematic region of the process. The most interesting kinematics at hight energies
 is the forward one. Because of that we investigate the photon-photon scattering

\begin{equation}\label{gamma}
\gamma^* (p)~ \gamma^* (q) \to  \gamma^* (p')~ \gamma^*(q'),
\end{equation}
with all photons being off-shell, via a single quark loop. We consider this reaction in the forward kinematics

\begin{equation}\label{fkin}
s = (p + q)^2 \gg -t = - (p'- p)^2 .
\end{equation}

 In order to be in agreement with the conventional notations, we
denote the photon virtualities as follows:

\begin{equation}\label{virt}
|p^2| = Q^2_1, |p'^2| = Q'^2, |q^2| = Q^2_2, |q'^2| = Q'^2_2,
\end{equation}
so that $Q^2_{1,2}, Q'^2_{1,2}$ are positive. We presume that $Q^2_{1,2} \approx Q'^2_{1,2}$.
In what follows we consider the case when $s \gg Q^2_{1,2}, Q'^2_{1,2}$, i.e.
when
\begin{equation}\label{wq}
s \gg |t|, Q^2_{1,2}.
\end{equation}
In contrast, we do not fix any hierarchy between $Q^2_{1,2}$ and $t$ and consider all possible situations.
Then, throughout the paper we will focus on the unpolarized  initial and final photons.
We will calculate the amplitude $A_{\gamma\gamma}$ of the reaction (\ref{gamma}) in the Double-Logarithmic
Approximation (DLA).
The imaginary part (with respect to $s$) of this amplitude was calculated in Ref.~\cite{bl} in the collinear
kinematics, i.e. in the kinematics (\ref{fkin}) with $t =0$, and
under the approximation of fixed QCD coupling $\alpha_s$. We check and confirm the results obtained in
Ref.~\cite{bl} and,
in contrast, we account for the running $\alpha_s$ effects, using
the results of Ref.~\cite{egtquark}.
 In our approach $\alpha_s$ runs in every rung of each involved
Feynman ladder graph. Then we consider the process (\ref{gamma}) in the forward kinematics, with $t \neq 0$ and
obtain a complete expression for the amplitude of this process.

In our calculations we compose and solve Infra-Red Evolution
Equations (IREE) for $A_{\gamma\gamma}$. The key point of the IREE method is the property of factorization of the double logarithmic (DL)
 contributions of the softest partons (i.e. the partons with minimal transverse momenta) out
of the scattering amplitudes. This remarkable property of the softest photons was first proved by
V.N.~Gribov\cite{g} in the QED context and then its generalization to the non-Abelian theories was
obtained  in
Ref.~\cite{l} and \cite{kl}, where the IREE method was suggested  to calculate
in DLA amplitudes of quark-antiquark scattering.
After that, the IREE method proved to be a simple and effective method to calculate in DLA
amplitudes of various inclusive
and exclusive processes in QCD and Standard Model, with both fixed and running $\alpha_s$, see e.g.
the overviews in Ref.~\cite{egtsum}.

The aim of our paper is to calculate the amplitude $M_{\gamma\gamma}$ of the process (\ref{gamma}) in the forward kinematics (\ref{fkin})
with non-zero value of $t$ and arbitray relations between $t$ and $Q^2_{1,2}$.
Throughout the paper we deal with running $\alpha_s$. Technology of composing IREE involves matching $M_{\gamma\gamma}$
with amplitude $A_{\gamma\gamma}$ calculated in the collinear kinematics. Following this pattern,
we in the first place calculate amplitude
$A_{\gamma\gamma}$ in the collinear kinematics, examining the cases of running and fixed $\alpha_s$
thereby checking results of Ref.~\cite{bl}, and then  proceed to calculating $M_{\gamma\gamma}$ in
the region of non-zero $t$.

Our paper is organized as follows: In Sects.~II-V we consider the
photon-photon scattering in collinear kinematics, i.e. in kinematics (\ref{fkin}) with $t = 0$.
In Sect.~II we briefly mention the lowest-order results for $A_{\gamma\gamma}$.
In Sect.~III we compose and solve IREE for $A_{\gamma\gamma}$, expressing it terms of auxiliary amplitudes
describing photon-quark scattering. In Sect.~IV we compose and solve IREE for the auxiliary amplitudes and
express them in terms of amplitudes of the quark-antiquark annihilation in the forward kinematics.
Using the obtained results, in Sect.~V we express the photon-photon scattering amplitudes
through the quark-quark amplitude.
Then in Sect.~VI we use results of Sect.~V in order to calculate the photon-photon scattering amplitude
$M_{\gamma\gamma}$ in kinematics (\ref{fkin}) at $t \neq 0$.
In Sect.~VII we discuss the results obtained in Sect.~V, VI. Here we consider the
high-energy asymptotics of $A_{\gamma\gamma}$ and compare them to the parent amplitudes, thereby
defining the applicability region for non-vacuum Reggeons.
 Finally, Sect.~VIII is for our concluding
remarks.

\section{Lowest-order amplitudes in the collinear kinematics}

First of all we consider the "Born", i.e. the simplest, case, where only quark box diagrams contribute.
We also suggest that $t \approx 0$. In this case
the amplitude $A_{B}$ of the process (\ref{gamma}) in the lowest order, with the quark masses neglected,
 consists of two terms:

\begin{equation}\label{ab}
A_B = B + B',
\end{equation}
where
\begin{eqnarray}\label{aborn}
B =\imath \frac{e^4}{(2 \pi)^4} \int d^4 k \frac{Tr\left[\gamma_{\nu}\left(\hat{q} + \hat{k}\right)
\gamma_{\mu}\hat{k}\gamma_{\lambda}\left(\hat{k}- \hat{p}\right)\gamma_{\rho}
\hat{k}\right]}
{k^2 k^2\left(q + k\right)^2 \left(k - p\right)^2}
\\ \nonumber
l_{\mu} (q)l_{\lambda} (p) l^*_{\nu}(q') l^*_{\lambda} (p')
\end{eqnarray}
and $B'$ can be obtained from (\ref{aborn}) by replacing $q \to - q$. The important property of (*** amplitude ***) $B$ is
that $\Im_s B \neq 0$ whereas $\Im_s B' = 0$. By this reason we will not consider $B'$ and focus on $B$ only.
In Eq.~(\ref{aborn}) we have neglected the quark mass and introduced the following notations:
$k$ is the loop momentum, $l_{\mu}(q), l_{\lambda}(p)$ are the polarization
vectors of the incoming photons and $l^*_{\nu}(q), l^*_{\rho}(p)$ stand for the polarization vectors of
the outgoing photons. Throughout the present paper we consider the case of the unpolarized photons  and
use for them the Feynman gauge
where the averaging over the photon polarizations can be done using the following replacements:

\begin{equation}\label{densmatr}
l_{\mu}(q) l^*_{\nu}(q) = -g_{\mu\nu}/2,~~l_{\lambda}(p)l^*_{\rho}(p) = -g_{\lambda\rho}/2
\end{equation}

and therefore

\begin{eqnarray}\label{tcol}
Tr\left[\gamma_{\nu}\left(\hat{q} + \hat{k}\right)
\gamma_{\mu}\hat{k}\gamma_{\lambda}\left(\hat{k}- \hat{p}\right)\gamma_{\rho}
\left(\hat{k} + \hat{p}' - \hat{p}\right)\right]
l_{\mu} (q)l_{\lambda} (p) l^*_{\nu}(q') l^*_{\lambda} (p')
\\ \nonumber
= Tr\left[\left(\hat{q} + \hat{k}\right)
\hat{k}\left(\hat{k}- \hat{p}\right)
\left(\hat{k} + \hat{p}' - \hat{p}\right)\right]
\\ \nonumber
\approx - Tr\left[\hat{q}
\hat{k}\hat{p}\hat{k}\right]
=  2 \left[w k^2 - 2pk 2qk\right],
\end{eqnarray}
where we have used the standard notation $w = 2pq$. For the next step, it is convenient to introduce the Sudakov
representation\cite{sud} for the soft momentum $k$:

\begin{equation}\label{sud}
k = - \alpha \widetilde{q} + \beta \widetilde{p} + k_{\perp},
\end{equation}
where the light-cone momenta $\widetilde{p},\widetilde{q}$ are made of the photon momenta $p$ and $q$:
\begin{equation}\label{pq}
\widetilde{p} = p - x_p q,~ \widetilde{q} = q - x_q p,~ x_p \approx Q^2_1/w,~x_q \approx Q^2_2/w,
\end{equation}
so that

\begin{equation}\label{sudinv}
2pk = - \alpha w + \beta x_p w,~~2qk = \beta w - \alpha x_q w,~
k^2 = - \alpha\beta w - k^2_{\perp}.
\end{equation}

In terms of the Sudakov variables Eq.~(\ref{tcol}) looks much simpler:

\begin{equation}\label{tcolsud}
2 \left[w k^2 - 2pk 2qk\right] \approx - 2 w k^2_{\perp} .
\end{equation}
Corrections to Eq.~(\ref{tcolsud}) are $\sim p^2, q^2$. Accounting for them is beyond the DLA accuracy, so we drop them.

Therefore, the DL contribution to
amplitude $B$ of Eq.~(\ref{aborn})
in collinear kinematics is
given by the following expression of the Sudakov type:

\begin{eqnarray}\label{abornsud}
B &=& - \imath \frac{e^4}{16 \pi^3}\int
\frac{d \alpha d \beta d k^2_{\perp}  w^2 k^2_{\perp}}
{k^2 k^2\left(x_q w + \beta w - \alpha x_q w + k^2\right) \left(x_p w + \alpha w - \beta x_p w + k^2 \right)}
\\ \nonumber
&\approx&  \imath \frac{e^4}{16 \pi^3}\int
\frac{d \alpha d \beta d k^2_{\perp}  w^2 }
{k^2\left(x_q w + \beta w + k^2\right) \left(x_p w + \alpha w + k^2 \right)}.
\end{eqnarray}

We have used in (\ref{abornsud}) that in DLA the integrand does not depend on the
azimuthal angle.

\subsection{Massless external photons}

This case is the simplest. Here $p^2 = q^2 = 0$, so the on-shell Born amplitude $B_{on}$ is

\begin{eqnarray}\label{aborndl}
B_{on} &=&  \imath \frac{e^4}{16 \pi^3}\int
\frac{d \alpha d \beta d k^2_{\perp}  w^2 }
{k^2\left(\beta w + k^2\right) \left( \alpha w + k^2 \right)}
\\ \nonumber
&=& -\frac{e^4}{8 \pi^2}\int_0^1 d \beta \int_0^s d k^2_{\perp}  \frac{w }
{k^2_{\perp}\left(w\beta  - k^2_{\perp}\right)}
= -\frac{e^4}{16 \pi^2} \ln^2 (w/\mu^2) \approx -\frac{e^4}{16 \pi^2} \ln^2 (s/\mu^2),
\end{eqnarray}
where we have introduced the infrared (IR) cut-off $\mu$ in the transverse space:
$k^2_{\perp} \gg \mu^2$.
In order to use the cut-off and at the same time neglect the quark masses, $\mu$ should obey the inequality
$\mu \gg m_{quark}$.
 With our accuracy, we have neglected the difference between $s$ and $w$ in Eq.~(\ref{aborndl}) and will do so throughout the paper.
%

\subsection{Off-shell external photons}

Here we consider the case of the off-shell photons. After integrating $B$ of Eq.~(\ref{abornsud}) over $k_{\perp}$, we arrive at

\begin{equation}\label{abornab}
B = -  \frac{e^4}{8 \pi^2}\int_0^1 d \alpha
\int_0^1 d \beta
\frac{\Theta (\alpha\beta - \lambda)}
{\left(\beta + x_q \right) \left(\alpha + x_p  \right)},
\end{equation}
where $\lambda = \mu^2/s$. Depending on the ratio between the photon virtualities and $\mu^2$, there are two different cases:\\

\textbf{(a) Moderately Virtual photons} \\

We call so the case, when virtualities $Q^2_1$ and $Q^2_2$ are sizable but not too great and obey the inequality

\begin{equation}\label{smallq}
Q^2_1 Q^2_2 \ll s \mu^2.
\end{equation}

\begin{figure}[h]
  \includegraphics[width=.4\textwidth]{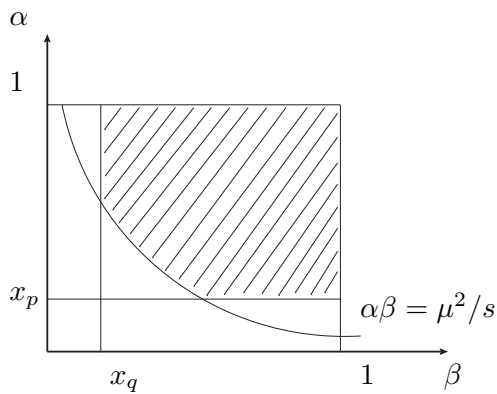}
  \caption{\label{gammafig1} Integration region for Moderately Virtual photons.}
\end{figure}

The integration region in this case is depicted in Fig.~\ref{gammafig1}

and therefore the off-shell Born amplitude $B_{\gamma\gamma}^{(M)}$ in the kinematics (\ref{smallq}) is

\begin{eqnarray}\label{born1}
B_{\gamma\gamma}^{(M)} &=& -  \frac{e^4}{8 \pi^2}\int_{x_p}^1 \frac{d \alpha}{\alpha}
\int_{x_q}^1 \frac{d \beta}{\beta} \Theta (\alpha\beta - \lambda)
\\  \nonumber
 &=& -\frac{e^4}{16 \pi^2} \left[\ln^2 (s/\mu^2)- \ln^2 (p^2/\mu^2) - \ln^2 (q^2/\mu^2)\right].
\end{eqnarray}

%

\textbf{(b) Deeply Virtual photons}\\

\begin{figure}[h]
  \includegraphics[width=.4\textwidth]{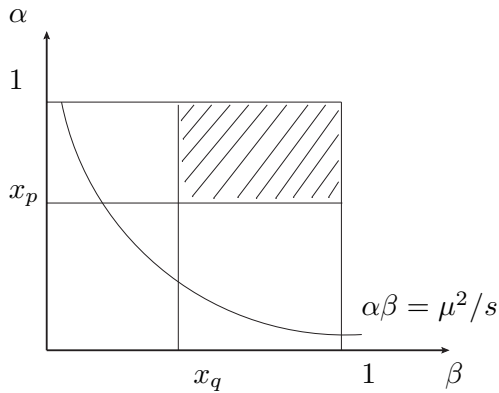}
  \caption{\label{gammafig2} Integration region for Deeply Virtual photons.}
\end{figure}

 On the contrary when the photon virtualities are so great that

\begin{equation}\label{bigq}
Q^2_1Q^2_2 \gg s \mu^2,
\end{equation}

the integration region does not include or touch the line $s \alpha\beta = \mu^2$ (see Fig.~\ref{gammafig2}),

 so
the amplitude $B_{\gamma\gamma}^{(M)}$  does not depend on $\mu$ and becomes IR stable:

\begin{equation}\label{born2}
B_{\gamma\gamma}^{(D)} = -  \frac{e^4}{8 \pi^2}\int_{x_p}^1 \frac{d \alpha}{\alpha}
\int_{x_q}^1 \frac{d \beta}{\beta} = -\frac{e^4}{8 \pi^2} \ln (s/Q^2_1)\ln(s/Q^2_2).
\end{equation}


\section{Photon-photon amplitudes in DLA}

In this Sect. we account for DL corrections to the Born amplitudes $B^{M}_{\gamma\gamma},~B^{D}_{\gamma\gamma} $ and express the amplitude $A_{\gamma \gamma}(s,Q^2_1, Q^2_2)$ of the process (\ref{gamma}) at
$t = 0$. We do it with constructing and solving IREE for $A_{\gamma \gamma}(s,Q^2_1, Q^2_2)$.
As a result, we represent $A_{\gamma \gamma}(s,Q^2_1, Q^2_2)$ in terms of auxiliary amplitudes $A_{\gamma q}$ and
$A_{q \gamma}$  that correspond respectively to the $t-$ channel annihilation of the pair of photons into quarks

\begin{equation}\label{gammaq}
\gamma^* (p)  \gamma^* (q) \to q (p'_1) \bar{q} (p'_2),
\end{equation}
and to the inverse process.
According to the IREE technology, we start with introducing a cut-off $\mu$ in the transverse space:

\begin{equation}\label{mu}
k_{\perp} \gg \mu,
\end{equation}
where $k_{\perp }$ refers to the transverse momenta of virtual quarks or gluons. In order to handle virtual quarks and gluons
equally, we choose $\mu$ much greater than the masses of involved quarks, which allows us to neglect the quark masses.
After that, the amplitude
 $A_{\gamma \gamma}$ becomes $\mu$-dependent, so we can evolve it with respect to $\mu$ and thereby compose IREE
 for $A_{\gamma \gamma}$. It is convenient to deal with $A_{\gamma \gamma}$ through its Mellin transformation
 $F_{\gamma \gamma}$. We will use the Mellin transform as follows:

 \begin{equation}\label{mellin}
A_{\gamma \gamma}(s, Q^2_1, Q^2_2) = \int_{- \imath \infty}^{\imath \infty}
\frac{d \omega}{2 \pi \imath} \left(s/\mu^2\right)^{\omega} F_{\gamma \gamma}(\omega, Q^2_1,Q^2_2)
= \int_{- \imath \infty}^{\imath \infty}
\frac{d \omega}{2 \pi \imath} e^{\omega \rho} F_{\gamma \gamma}(\omega, y_1,y_2),
\end{equation}
where we have denoted
\begin{equation}\label{y12}
\rho = \ln (s/\mu^2),~~y_1 = \ln (Q^2_1/\mu^2),~~y_2 = \ln (Q^2_2/\mu^2).
 \end{equation}
We will address $F_{\gamma \gamma}(\omega, y_2,y_1)$ and $\left(s/\mu^2\right)^{\omega}$ as the Mellin amplitude and
the Mellin factor respectively. We would like to remind that in the context of the Regge processes
the Mellin transform in Eq.~(\ref{mellin}) is actually the asymptotic form of the Sommerfeld-Watson
representation for the positive signature amplitudes.
%
%
Before composing IREE for objects with several $\mu$-dependent variables like $A_{\gamma \gamma}(\rho, y_1, y_2)$, we
should order these variables.
We use the ordering of  Eq.~(\ref{wq}), complementing it by the restriction
$Q^2_1 \gg Q^2_2$ and arriving thereby at

\begin{equation}\label{rhoy}
\rho > y_1 > y_2.
\end{equation}
When we obtain expressions for $A_{\gamma \gamma}(\rho, y_1, y_2)$ under the ordering (\ref{rhoy}), we
will generalize our results on the case of the opposite ordering $y_1 < y_2$ and for $y_1 = y_2$ as well.
The general strategy of composing IREE prescribes
to start with considering the simplest case: we first compose the IREE for the on-shell amplitude $A_{\gamma \gamma}^{on}$, which
describes the process (\ref{gamma}) at $y_1 = y_2 = 0$ and
therefore depends on the largest variable $\rho$ only.  When $A_{\gamma \gamma}^{on}$ is found, we do next step,
considering the more involved case of amplitude $\widetilde{A}_{\gamma\gamma} (\rho, y_1)$ of the same process in
the kinematics $\rho > y_1 > y_2 = 0$. In order to specify a general solution of
the IREE for $\widetilde{A}_{\gamma\gamma} (\rho, y_1)$, we will use matching with
the on-shell amplitude $A_{\gamma \gamma}^{on}$, which has been found on the previous step. Then we repeat
the same to specify a general solution to the IREE for $A_{\gamma\gamma} (\rho, y_1, y_2)$.
Obviously, such procedure can be repeated as many times as one needs, allowing to
describe processes with arbitrary number of external kinematic invariants. We suppose that the amplitudes
$A_{\gamma \gamma}^{on} (\rho), \widetilde{A}_{\gamma\gamma} (\rho, y_1), A_{\gamma\gamma} (\rho, y_1, y_2)$
are related to the conjugated Mellin amplitudes $f_{\gamma \gamma}(\omega), \widetilde{F}_{\gamma\gamma} (\omega, y_1),
F_{\gamma\gamma} (\omega, y_1, y_2)$ by the Mellin transform (\ref{mellin}).

\begin{figure}[h]
\centerline{\includegraphics[width=.4\textwidth]{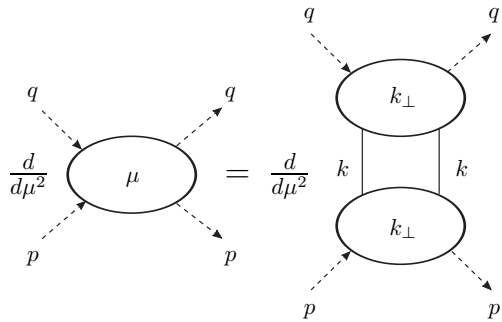}}
\caption{\label{gammafig3} Infra-Red Evolution Equation for the
amplitude $A_{\gamma\gamma}$. The dashed lines denote the external photons,
whereas the straight lines correspond to quarks. The blobs stand for
amplitudes calculated in DLA. The letters on the blobs denote the IR cut-offs
for the involved amplitudes.}
\end{figure}

Now we have got all set to compose IREE for the amplitudes of the process (\ref{gamma}).
The generic form of IREE for $A_{\gamma \gamma}$ is depicted in Fig.~\ref{gammafig3}.
Throughout this paper we will write the IREE directly in the $\omega$-space.

\subsection{All photons are nearly on-shell}

We start with calculation of $A_{\gamma \gamma}^{on}$
in the simplest kinematics where $Q^2_1 \approx Q^2_2 \lesssim \mu^2$.
We denote $f_{\gamma \gamma}(\omega)$ the Mellin amplitude for the photon-photon scattering
in the case when virtualities $Q^2_{1,2}$ are neglected, i.e. when

\begin{equation}\label{q12on}
y_2 = y_1 = 0.
\end{equation}

The IREE for $f_{\gamma \gamma}(\omega)$ is very simple. It represents  $f_{\gamma \gamma}(\omega)$
through two auxiliary Mellin amplitudes:

\begin{equation}\label{fon}
\omega f_{\gamma \gamma}(\omega) = \frac{1}{8 \pi^2} f_{\gamma q}(\omega) f_{q \gamma}(\omega),
\end{equation}
where $f_{\gamma q}(\omega)$ and $f_{q \gamma}(\omega)$ corresponds to the processes
(\ref{gammaq}) and the reversal process respectively. In fact, $f_{\gamma q}(\omega) = f_{q \gamma}(\omega)$.

\subsection{One of the photons is on-shell and the other is off-shell }

Let us consider the more complicated case when

\begin{equation}\label{y2mu}
\rho > y_1 > y_2 = 0,
\end{equation}
i.e. $Q^2_1 \gg Q^2_2 \sim \mu^2$,
and denote $\widetilde{F}_{\gamma\gamma} (\omega, y_1)$ the amplitude corresponding to that case. It obeys the
following IREE:

\begin{equation}\label{eq2gen}
\frac{\partial \widetilde{F}_{\gamma\gamma}}{\partial y_1}
+ \omega \widetilde{F}_{\gamma\gamma} = \frac{1}{8 \pi^2} F_{\gamma q} (\omega, y_1)
f_{q \gamma} (\omega).
\end{equation}
where amplitudes $F_{\gamma q} (\omega, y_1)$ and $f_{q \gamma} (\omega)$ are supposed to be
calculated independently.
Once they are known, the general solution to Eq.~(\ref{eq2gen}) is

\begin{equation}\label{sol2gen}
 \widetilde{F}_{\gamma\gamma} = e^{-\omega y_1}
\left[C_2 (\omega) + \frac{1}{8 \pi^2}  f_{q \gamma} (\omega) \int_0^{y_1} d y' e^{\omega y'} F_{\gamma q} (\omega, y')  \right] \ .
\end{equation}

In order to specify an unknown function $C_2$ in Eq.~(\ref{sol2gen}) we use the matching:

\begin{equation}\label{match2}
\widetilde{F}_{\gamma\gamma} (\omega, y_1)|_{y_1 = 0} = f_{\gamma\gamma} (\omega),
\end{equation}
where $f_{\gamma\gamma} (\omega)$ is defined in Eq.~(\ref{fon}). Therefore,
$\widetilde{F}_{\gamma\gamma} (\omega, y_1)$ in kinematics (\ref{y2mu}) is represented in terms of the photon-quark amplitudes:

\begin{equation}\label{sol2}
 \widetilde{F}_{\gamma\gamma} (\omega, y_1) = e^{-\omega y_1}\left[ \frac{1}{8 \pi^2 \omega}  f_{\gamma q} (\omega)
 f_{q \gamma} (\omega) + \frac{1}{8 \pi^2}
 f_{q \gamma} (\omega) \int_0^{y_1} d y e^{\omega y} F_{\gamma q} (\omega, y)\right],
\end{equation}
where $F_{\gamma q} (\omega, y)$ is the photon-quark amplitude at $y \neq 0$.

\subsection{Off-shell photons with moderate  virtualities}

We call the moderately virtual kinematics the case when
$Q^2_1 \gg \mu^2$ and $Q^2_2 \gg \mu^2$ but $Q^2_1 Q^2_2 \ll s \mu^2$. In the logarithmic
variables it means that

\begin{equation}\label{modoffshell}
\rho > y_2 + y_1.
\end{equation}

The IREE for the amplitude $F_{\gamma\gamma}^{(M)} (\omega, y_1, y_2)$  in the kinematic region (\ref{modoffshell}) is

\begin{equation}\label{eq12gen}
\frac{\partial F_{\gamma\gamma}^{(M)}}{\partial y_2} + \frac{\partial F_{\gamma\gamma}^{(M)}}{\partial y_1}
+ \omega F_{\gamma\gamma}^{(M)} = \frac{1}{8 \pi^2} F_{\gamma q} (\omega, y_1) F_{q \gamma} (\omega, y_2).
\end{equation}

In order to use the symmetry with respect to $y_1, y_2$ of the differential operator
in (\ref{eq12gen}) and simplify the IREE, we have introduced new variables $\xi, \eta$:

\begin{equation}\label{xieta}
\xi = y_1 + y_2,~~\eta = y_1 - y_2.
\end{equation}

%

Eq.~(\ref{eq12gen}) in terms of $\xi, \eta$ takes a simpler form:

\begin{equation}\label{eq12xieta}
2 \frac{\partial F_{\gamma\gamma}^{(M)}}{\partial \xi}
+ \omega F_{\gamma\gamma}^{(M)} = \frac{1}{8 \pi^2} F_{\gamma q} \left(\omega, y_1 \right)
F_{q \gamma} \left(\omega, y_2\right).
\end{equation}

A general solution to Eq.~(\ref{eq12xieta})
is

\begin{equation}\label{sol12gen}
F_{\gamma\gamma}^{(M)} = e^{- \omega \xi/2} \left[C(\omega, \eta) + \frac{1}{16 \pi^2} \int_0^{\xi} d \xi'
e^{\omega \xi'/2} F_{\gamma q} (\omega, y'_1) F_{q \gamma} (\omega, y'_2) \right],
\end{equation}
with $C(\omega, \eta)$ being an arbitrary function and the variables $y'_1,y'_2$ are defined as
follows:

\begin{equation}\label{y12prime}
y'_1 = (\xi' + \eta)/2,~~y'_2 = (\xi' -\eta)/2).
\end{equation}

In order to specify $C(\omega, \eta)$, we use the matching of $F_{\gamma\gamma}^{(M)} (\omega, y_1,y_2)$
with an amplitude $\widetilde{F}_{\gamma\gamma} (\omega, y_1)$
of the same process but in the simpler kinematic regime (\ref{y2mu}) considered above:

\begin{equation}\label{match1}
F_{\gamma\gamma}^{(M)} (\omega, y_1,y_2)|_{y_2 = 0} = \widetilde{F}_{\gamma\gamma} (\omega, y_1),
\end{equation}
where amplitude $\widetilde{F}_{\gamma\gamma} (\omega, y_1)$ is given by Eq.~(\ref{sol2}).
Combining Eqs.~(\ref{match1}), (\ref{sol12gen}) and (\ref{sol2}), we arrive at the following expression for
$F_{\gamma\gamma}$:

\begin{eqnarray}\label{fm}
F_{\gamma\gamma}^{(M)} &=& e^{- \omega \xi/2} \left[e^{\omega \eta/2}
\widetilde{F}_{\gamma\gamma} (\omega, \eta)
+ \frac{1}{16 \pi^2}  \int_{\eta}^{\xi} d \xi' e^{\omega \xi'/2}
F_{\gamma q} (\omega, y'_1) F_{q \gamma} (\omega, y'_2)\right].
\end{eqnarray}

Substituting Eq.~(\ref{fm}) in (\ref{mellin}), we arrive at the expression for
the amplitude $A_{\gamma\gamma}^{(M)}$ at moderate virtualities $Q^2_{1,2}$:

\begin{eqnarray}\label{am}
A_{\gamma\gamma}^{(M)} &=&
\int_{- \imath \infty}^{\imath \infty}
\frac{d \omega}{2 \pi \imath}e^{\omega (\rho- \xi/2)}
\left[e^{\omega \eta/2}
\widetilde{F}_{\gamma\gamma} (\omega, \eta)
+ \frac{1}{16 \pi^2}  \int_{\eta}^{\xi} d \xi' e^{\omega \xi'/2}
F_{\gamma q} (\omega, y'_2) F_{q \gamma} (\omega, y'_1)\right]
\\ \nonumber
&=& \int_{- \imath \infty}^{\imath \infty}
\frac{d \omega}{2 \pi \imath}
\left(\frac{s}{\sqrt{Q^2_1 Q^2_2}}\right)^{\omega}
\left[e^{\omega \eta/2}
\widetilde{F}_{\gamma\gamma} (\omega, \eta)
+ \frac{1}{16 \pi^2}  \int_{\eta}^{\xi} d \xi' e^{\omega \xi'/2}
F_{\gamma q} (\omega, y'_2) F_{q \gamma} (\omega, y'_1)\right],
\end{eqnarray}
where $\widetilde{F}_{\gamma\gamma}$ is expressed in Eq.~(\ref{sol2}) through
the auxiliary amplitudes. Eqs.~(\ref{fm},\ref{am}) are obtained under the assumption
of Eq.~(\ref{modoffshell}) that $y_1 > y_2$. Writing $A_{\gamma\gamma}^{(M)}$ and
$F_{\gamma\gamma}^{(M)}$ in terms of variables $\zeta,\eta$ makes easy to see
that the reverse assumption $y_1 > y_2$ leads to the expressions for
$A_{\gamma\gamma}^{(M)}, F_{\gamma\gamma}^{(M)}$, with $\eta$ replaced by
$- \eta$. Therefore, replacing $\eta$ by $|\eta|$ in Eqs.~~(\ref{fm},\ref{am})
allows us to embrace the both cases. After the replacement has been done, Eqs.~~(\ref{fm},\ref{am}) are
indeed invariant to the exchange $y_1 \leftrightarrows y_2$.

\subsection{Deeply-virtual photons}

When $Q^2_1 >> \mu^2$ and $Q^2_2 >> \mu^2$, and their product is also great, $Q^2_1 Q^2_2 >> s \mu^2$,
the inequality in Eq.~(\ref{modoffshell})
is replaced by the opposite one

\begin{equation}\label{deepy}
\rho < y_2 + y_1.
\end{equation}

We address such photons as deeply-virtual ones. The principal difference between this case and the case of moderately-virtual
photons is that the scattering amplitude $A_{\gamma\gamma}^{(D)} (\rho,y_1,y_2)$ in the kinematics (\ref{deepy})
does not depend on $\mu$, so the IREE for it is very simple:

\begin{equation}\label{eqad}
\frac{\partial A_{\gamma\gamma}^{(D)}}{\partial \rho}
+ \frac{\partial A_{\gamma\gamma}^{(D)}}{\partial y_1 } +
\frac{\partial A_{\gamma\gamma}^{(D)}}{\partial y_2} = 0.
\end{equation}

A general solution to Eq.~(\ref{eqad}) can be written in different ways. The most convenient
way for our goal is

\begin{equation}\label{adeepgen}
A_{\gamma\gamma}^{(D)} = M (\rho - y_1, \rho - y_2),
\end{equation}
with $M$ being an arbitrary analytic function. In order to specify $M$ we
use the matching with the amplitude $A_{\gamma\gamma}^{(M)}$ of the same process but in the region
(\ref{modoffshell}). It means that

\begin{equation}\label{match12}
A_{\gamma\gamma}^{(D)} (\rho-y_1, \rho - y_2)|_{\rho = y_1 + y_2}
= A_{\gamma\gamma}^{(M)}(\rho,y_1,y_2)|_{\rho = y_1 + y_2}
\equiv \widetilde{A}_{\gamma\gamma}^{(M)}(y_1,y_2).
\end{equation}
Replacing $y_1 \to \rho - y_2$ and  $y_2 \to \rho - y_1$  in $\widetilde{A}_{\gamma\gamma}^{(M)}(y_1,y_2)$
immediately allows us to express $A_{\gamma\gamma}^{(D)}$
through $\widetilde{A}_{\gamma\gamma}^{(M)}$ in the whole the region $\rho \leq y_1 + y_2$:

\begin{equation}\label{ad}
A_{\gamma\gamma}^{(D)} (\rho,y_1,y_2) = \widetilde{A}_{\gamma\gamma}^{(M)}(\rho - y_2, \rho - y_1),
\end{equation}
or, in terms of the Mellin transform,

\begin{equation}\label{afd}
A_{\gamma\gamma}^{(D)} =
\int_{- \imath \infty}^{\imath \infty}
\frac{d \omega}{2 \pi \imath}e^{\omega (\rho- \xi/2)}
\left[e^{\omega \eta/2}
\widetilde{F}_{\gamma\gamma} (\omega, \eta)
+ \frac{1}{16 \pi^2}  \int_{\eta}^{2 \rho -\xi} d \xi' e^{\omega \xi'/2}
F_{\gamma q} (\omega, y'_2) F_{q \gamma} (\omega, y'_1)\right] \ .
\end{equation}

Eq.~(\ref{afd}) demonstrate that, in contrast to the previous cases, the variable $\rho$
participates not only in the Mallin factor but also in the expression in parentheses.
This should be taken as a clear warning not to use the Mellin amplitudes
for the matching. Indeed, applying the Mellin transform to Eq.~(\ref{eqad}) converts it into
the following equation for the Mellin amplitude $F_{\gamma\gamma}^{(D)}$:

\begin{equation}\label{eqfd}
\omega F_{\gamma\gamma}^{(D)}
+ \frac{\partial F_{\gamma\gamma}^{(D)}}{\partial y_1 } +
\frac{\partial F_{\gamma\gamma}^{(D)}}{\partial y_2} = 0,
\end{equation}
(*** comment: in the above equation I removes extra $+$ ***) or

\begin{equation}\label{eqfdksi}
\omega F_{\gamma\gamma}^{(D)}  +
\frac{\partial F_{\gamma\gamma}^{(D)}}{\partial \xi} = 0,
\end{equation}
with the obvious general solution:

\begin{equation}\label{fdgen}
F_{\gamma\gamma}^{(D)} = \Phi (\omega, \eta) e^{- \omega \xi},
\end{equation}
where an unspecified function $\Phi$ is supposed to be found through matching with $F_{\gamma\gamma}^{(M)}$ of Eq.~(\ref{fm}) at $\rho = \xi$.
However, it cannot be done because $\Phi$ by definition does not depend on $\xi$ whereas $F_{\gamma\gamma}^{(M)}$ depends on it.
So, the matching can be done for the amplitudes $A_{\gamma\gamma}^{(D)}, A_{\gamma\gamma}^{(M)}$.
We consider this issue in more detail in Sect.~V.

\section{Auxiliary amplitudes}

In the previous Sect. we obtained amplitudes $A_{\gamma\gamma}^{(M,D)}$
in terms of auxiliary amplitudes $A_{\gamma q},A_{q \gamma}$.
corresponding to the process of Eq.~(\ref{gammaq}) and the inverse process respectively.
We denote $F_{\gamma q}(\omega,y)$ and $F_{q \gamma}(\omega,y)$ the Mellin amplitudes  related to
$A_{\gamma q},A_{q \gamma}$   respectively. We remind that throughout the paper we neglect the quark masses.
We will compose and solve IREE for them, considering first the simplest kinematics,
where the photons are on-shell and then move to the case of off-shell photons.
As $A_{\gamma q}$ and $A_{q \gamma}$ are much alike, we consider in detail dealing with
$F_{\gamma q}$ only.

\subsection{Photon-quark amplitude with on-shell photon}

\begin{figure}[h]
\centerline{\includegraphics[width=.6\textwidth]{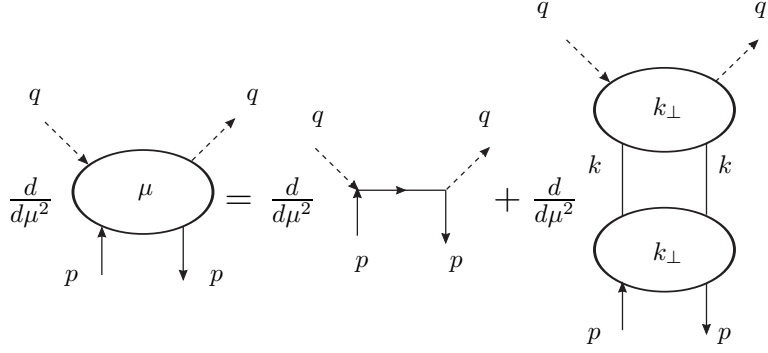}}
  \caption{\label{gammafig4} Infra-Red Evolution Equation for the amplitude $A_{\gamma q}$.}
\end{figure}

We consider the case when $y = 0$ and denote $f_{\gamma q}(\omega)$
the Mellin amplitude of such a process.
The IREE for $f_{\gamma q} (\omega) $ is depicted in Fig.~\ref{gammafig4}.

In the $\omega$-space it is
\begin{equation}\label{eqgammaqtilde}
 f_{\gamma q}(\omega) =
\frac{a_{\gamma q}}{\omega} + \frac{1}{8 \pi^2 \omega}
f_{\gamma q} (\omega) f_0 (\omega),
\end{equation}
where $a_{\gamma q}/\omega$, with $a_{\gamma q} = e^2$, is the Born amplitude and
$f_0$ is the quark-quark amplitude. It includes  the
total resummation of DL contributions as well as accounts for the running coupling.
The solution to  Eq.~(\ref{eqgammaqtilde}) is

\begin{equation}\label{gammaqtilde}
f_{\gamma q}(\omega) = \frac{a_{\gamma q}}{\omega - H(\omega)},
\end{equation}
where, by convenience reason we have introduced the notation $H (\omega) = (1/8 \pi^2) f_0 (\omega)$. In DIS (*** process ***) $H$ plays the role of the non-singlet
anomalous dimension calculated in DLA.

\subsection{Off-shell photons}

IREE for $F_{\gamma q}(\omega, y)$ is

\begin{equation}\label{eqgq}
\frac{\partial}{\partial y}F_{\gamma q} (\omega, y) +  \omega F_{\gamma q}(\omega, y) =
\frac{1}{8 \pi^2}
F_{\gamma q} (\omega, y) f_0 (\omega).
\end{equation}

A general solution to Eq.~(\ref{eqgq}) is

\begin{equation}\label{{solgqgen}}
F_{\gamma q} (\omega, y) = C_{\gamma q}(\omega) e^{-y [\omega - H (\omega)]},
\end{equation}
with  $C_{\gamma g}$ being an arbitrary function.
To specify $C_{\gamma g}$ we use the matching

\begin{equation}\label{matchq}
F_{\gamma q} (\omega,y)|_{y = 0} = \tilde{F}_{\gamma q} (\omega),
 \end{equation}
with $\tilde{F}_{\gamma q} (\omega)$ defined in Eq.~(\ref{gammaqtilde}).
The use of Eq.~(\ref{matchq}) leads to
\begin{equation}\label{fgammag}
F_{\gamma q}(\omega, y) = \frac{a_{\gamma q}}{\omega - H(\omega)}
e^{-y [\omega - H (\omega)]}
\end{equation}

Now the auxiliary amplitude $F_{\gamma q} (\omega, y) f_0 (\omega)$ is expressed through
the on-shell quark-quark amplitude $f_0$ which is well-known. It was calculated in Ref.~\cite{kl},
with $\alpha_s$ being fixed.

\subsection{Quark-quark amplitude}

Amplitude $f_0$ was obtained in Ref.~\cite{kl}.
It satisfies the simple algebraic equation

\begin{equation}\label{eqfqq}
f_0 = \frac{a_0}{\omega} + \frac{1}{8 \pi^2 \omega} f_0 f_0,
\end{equation}
where $a_0/\omega$ is the Born amplitude. Solving Eq.~(\ref{fqq}), one arrives at
the explicit expression for $f_0$:

\begin{equation}\label{fqq}
f_0 = 4 \pi^2 \left[\omega - \sqrt{\omega^2 - a_0/(2 \pi^2)}\right].
\end{equation}
Eq.~(\ref{fqq}) is true for the both cases of fixed and running $\alpha_s$ but
$a_0$ in those cases are different. For fixed QCD coupling, $a_0 \equiv a_0^{fix}$
was obtained in Ref.~\cite{kl}:

\begin{equation}\label{aqqfix}
a_0^{fix} = 4 \pi \alpha_s^{fix} C_F,
\end{equation}
with $C_F = (N^2-1)/(2N) = 4/3$, while at running $\alpha_s$ it depends on $\omega$
(see \cite{egtquark} for detail):
\begin{equation}\label{aqqrun}
a_0(\omega) = \frac{4 \pi C_F}{ b} \left[\frac{\zeta}{\zeta^2 + \pi^2}
- \int_0^{\infty}\frac{d \rho e^{- \omega \rho}}
{(\rho + \zeta)^2 + \pi^2}\right] ,
\end{equation}
where $\zeta = \ln \left(\mu^2/\Lambda^2_{QCD}\right)$ and $b = (11 N - 2 n_f)/(12 \pi^2)$ is the
standard notation for first coefficient of the Gell-Mann - Low function.
%
%

%

\subsection{Representation of the auxiliary amplitudes through the quark amplitude}

Using Eq.~(\ref{eqfqq}) allows us to simplify
Eq.~(\ref{gammaqtilde}) for the auxiliary amplitude $f_{\gamma q}(\omega)$:

\begin{equation}\label{gammaqtilde1}
f_{\gamma q}(\omega) = \frac{a_{\gamma q}}{a_0} f_0 (\omega)
\end{equation}
as well as the expression for $F_{\gamma q} (\omega, y)$ in Eq.~(\ref{fgammag}):

\begin{equation}\label{gammaq1}
F_{\gamma q} (\omega, y) =
\frac{a_{\gamma q}}{a_0} f_0 (\omega)
e^{- y \left(\omega - H (\omega)\right)}.
\end{equation}

The only difference between IREE for $F_{\gamma q} (\omega, y)$ and $F_{q \gamma} (\omega, y)$
is in the use of different the factors $a_{\gamma q}$ and $a_{q \gamma}$ respectively, so
expressions for $\tilde{F}_{q \gamma} (\omega, y)$ and  $F_{q \gamma} (\omega, y)$ can be immediately obtained from
Eqs.~(\ref{gammaqtilde1}) and (\ref{gammaq1}):

\begin{equation}\label{qgammatilde1}
f_{q\gamma }(\omega) = \frac{a_{q\gamma }}{a_0} f_0 (\omega)
\end{equation}

\begin{equation}\label{qgamma1}
F_{q \gamma} (\omega, y) =
\frac{a_{q \gamma}}{a_0} f_0 (\omega)
e^{- y \left(\omega - H (\omega)\right)}.
\end{equation}

We define the factors $a_{\gamma q}$ and $a_{q \gamma}$ as follows:

\begin{equation}\label{agammaqborn}
a_{\gamma q} = e^2_q,~~a_{q \gamma} = -e^2_q  ,
\end{equation}
where $e_q$ is the electric charge of the loop quark.

\section{Representation of photon-photon amplitudes through quark-quark amplitudes}

\subsection{On-shell initial photons}

Substituting Eqs.~(\ref{qgammatilde1},\ref{gammaqtilde1}) in Eq.~(\ref{fon}) and using Eq.(\ref{eqfqq}), we obtain

\begin{equation}\label{fonqq}
f_{\gamma \gamma}(\omega) = \kappa
\left[f_0 (\omega) - \frac{a_0}{\omega} \right],
\end{equation}
with
\begin{equation}\label{kappa}
\kappa = \frac{a_{\gamma q} a_{q \gamma}}{a_0^2}.
\end{equation}
According to Eq.~(\ref{aqqrun}) $\kappa$ depends on $\omega$, when $\alpha_s$ is running,
so throughout the paper we will keep it under the Mellin integral sign.
The photon-photon scattering amplitude $A^{(on)}_{\gamma\gamma}$, all photons are on-shell, is

\begin{equation}\label{aon}
A^{(on)}_{\gamma\gamma} (s/\mu^2) =
\int_{-\imath \infty}^{\imath \infty} \frac{d \omega}{2 \pi \imath}e^{\omega \rho}~\kappa
 \left[f_0 (\omega) - \frac{a_0}{\omega} \right].
\end{equation}

\subsection{One of the photons is off-shell}

If $y_1 > y_2 = 0$, amplitude $\widetilde{F}_{\gamma \gamma}$ is given by Eq.~(\ref{sol2}) where
$\widetilde{F}_{\gamma\gamma}(\omega,y_1)$ is represented through the auxiliary amplitudes.
 Combining Eq.~(\ref{sol2},\ref{gammaq1}) and (\ref{fonqq}), we express $\widetilde{F}_{\gamma\gamma}(\omega,y_1)$
 in terms of $f_0(\omega)$:

\begin{equation}\label{fgammatildeqq}
\widetilde{F}_{\gamma\gamma}(\omega,y_1) = \kappa
e^{- \omega y_1} \left[
f_0 (\omega) e^{y_1H} - \frac{a_0}{\omega} \right].
\end{equation}

Therefore,

\begin{equation}\label{agammatildeqq}
\widetilde{A}_{\gamma\gamma}(\omega,y_1) =
\int_{-\imath \infty}^{\imath \infty} \frac{d \omega}{2 \pi \imath}  e^{\omega (\rho - y_1)}
~\kappa
\left[f_0 (\omega) e^{y_1 f_0 /(8 \pi^2)} - \frac{a_0}{\omega} \right].
\end{equation}

\subsection{Moderately-virtual photons}

Combining Eq.~(\ref{fm}) with Eqs.~(\ref{fgammatildeqq},\ref{gammaq1},\ref{qgamma1}) allows us to obtain
$F_{\gamma \gamma}^{(M)}$, so we can write the amplitude $A_{\gamma \gamma}^{(M)}$ in the MV region as follows:

\begin{equation}\label{amw}
A_{\gamma\gamma}^{(M)} =
\int_{- \imath \infty}^{\imath \infty} \frac{d \omega}{2 \pi \imath}
e^{\omega \rho} ~\kappa
 \left[ W_1 + W_2 \right]
,
\end{equation}
with

\begin{eqnarray}\label{wm}
W_1 &=& e^{-\omega (\xi +|\eta|)/2}
\left(- \frac{a_0}{\omega} + f_0 e^{|\eta| H}\right)
\\ \nonumber
W_2 &=& \left(f_0 - \frac{a_0}{\omega}\right)
\frac{\omega}{\sqrt{\omega^2 - a_0/(2 \pi^2)}} \left[e^{-(\xi + |\eta|)\omega/2 + \eta H} - e^{\xi (-\omega + H)}\right].
\end{eqnarray}

We remind
that the variables $\xi,\eta$ are defined in Eq.~(\ref{xieta}).
Eqs.~(\ref{amw},\ref{wm}) describe $A_{\gamma\gamma}^{(M)}$ at any ordering between
$y_1$ and $y_2$, i.e. at $y_1 > y_2$ and  $y_1 < y_2$; they also stand when $y_1 = y_2$.

\subsection{Deeply-virtual photons}

According to Eq.~(\ref{match12}), aamplitude $A_{\gamma\gamma}^{(M)}$ can be found  through matching with amplitude $A_{\gamma\gamma}^{(M)}$
at the border between the Deeply-Virtual and Moderately-Virtual regions, where

\begin{equation}\label{border}
\rho = \xi.
\end{equation}

As $\rho$ participates in the Mellin factor, the matching should involve the whole amplitudes
$A_{\gamma\gamma}^{(M)}, A_{\gamma\gamma}^{(D)}$ rather than $F_{\gamma\gamma}^{(M)}, F_{\gamma\gamma}^{(M)}$.
For performing the matching the easiest way, we replace Eq.~(\ref{amw}) by the following one:

 \begin{eqnarray}\label{amxi}
 A_{\gamma\gamma}^{(M)} =  \int_{- \imath \infty}^{\imath \infty} \frac{d \omega}{2 \pi \imath}
e^{(\omega \rho - \omega \xi/2)} ~\kappa F_1 (\omega, \eta)
-  \int_{- \imath \infty}^{\imath \infty} \frac{d \omega}{2 \pi \imath}
e^{(\omega \rho - \omega \xi + \xi H)}  \kappa F_2 (\omega)
,
 \end{eqnarray}
where

\begin{eqnarray}\label{f12}
F_1 (\omega, \eta) &=& e^{-\omega \eta/2}
\left[
\left(- \frac{a_0}{\omega} + f_0 e^{\eta H}\right)
+ \left(f_0 - \frac{a_0}{\omega}\right)
\frac{\omega}{\sqrt{\omega^2 - a_0/(2 \pi^2)}} e^{\eta H}
\right],
\\ \nonumber
F_2 (\omega) &=& \left(f_0 - \frac{a_0}{\omega}\right)
\frac{\omega}{\sqrt{\omega^2 - a_0/(2 \pi^2)}}.
\end{eqnarray}

Then at $\rho = \xi$ amplitude $A_{\gamma\gamma}^{(M)}$ becomes $\bar{A}_{\gamma\gamma}^{(M)}$:

\begin{equation}\label{abarxi}
\bar{A}_{\gamma\gamma}^{(M)} =  \int_{- \imath \infty}^{\imath \infty} \frac{d \omega}{2 \pi \imath}
e^{\omega \xi/2}~\kappa F_1 (\omega, \eta)
-  \int_{- \imath \infty}^{\imath \infty} \frac{d \omega}{2 \pi \imath}
e^{\xi H} ~\kappa F_2 (\omega)
\end{equation}
and therefore in the Deeply-Virtual region

\begin{equation}\label{adgen}
A_{\gamma\gamma}^{(D)} =  \int_{- \imath \infty}^{\imath \infty} \frac{d \omega}{2 \pi \imath}
e^{\omega (\rho - \xi/2)} ~\kappa F_1 (\omega, \eta)
-  \int_{- \imath \infty}^{\imath \infty} \frac{d \omega}{2 \pi \imath}
e^{ (2\rho - \xi) H} ~\kappa F_2 (\omega).
\end{equation}

The second integral in Eq.~(\ref{adgen}) can be dropped because it does not contain the standard Mellin factor
$e^{\omega \rho}$ (or $e^{\omega (2\rho- \xi)}$) which would prevent closing the
integration contour to the right, where the integrand does not
have singularities. Closing the contour to the right, we find that the integration over $\omega$ yields a zero.
So, we arrive at the following expression which is true for any ordering of $y_{1,2}$ and for the case $y_1 = y_2$:

\begin{eqnarray}\label{adsym}
A_{\gamma\gamma}^{(D)} &=&  \int_{- \imath \infty}^{\imath \infty} \frac{d \omega}{2 \pi \imath}
e^{\omega (\rho - \xi/2 - |\eta|/2)}
~\kappa \left[
\left(- \frac{a_0}{\omega} + f_0 e^{|\eta| H}\right)
+ \left(f_0 - \frac{a_0}{\omega}\right)
\frac{\omega}{\sqrt{\omega^2 - a_0/(2 \pi^2)}} e^{|\eta| H}
\right]
\\ \nonumber
&=&   \int_{- \imath \infty}^{\imath \infty} \frac{d \omega}{2 \pi \imath}
\left(\frac{s}{\sqrt{Q^2_1Q^2_2}}\right)^{\omega}
\left(\frac{Q^2_{\max}}{Q^2_{\min}}\right)^{\omega}
~\kappa \left[
\left(- \frac{a_0}{\omega} + f_0 e^{|\eta| H}\right)
+ \left(f_0 - \frac{a_0}{\omega}\right)
\frac{\omega}{\sqrt{\omega^2 - a_0/(2 \pi^2)}} e^{|\eta| H}
\right],
\end{eqnarray}
where we have denoted $Q^2_{\max} = \max [Q^2_{1,2}]$ and
$Q^2_{\min} = \min [Q^2_{1,2}]$.

The amplitudes $A_{\gamma\gamma}^{(M)}$ in Eq.~(\ref{amw}) and $A_{\gamma\gamma}^{(D)}$ in
Eqs.~(\ref{adsym},\ref{adq}) are represented in the
form different from the expressions for the same amplitudes obtained in Ref.~\cite{bl}, which is
unessential.
The main difference between our approach and Ref.~\cite{bl} is our accounting for the running
QCD coupling. In this case $a_0$ depends on $\omega$ (see Eq.~(\ref{aqqrun})).  Now let us remind that $A_{\gamma\gamma}^{(M)}$
and $A_{\gamma\gamma}^{(D)}$ are not  complete expressions for
amplitudes of the process (\ref{gamma}) in the collinear kinematics. In
order to account for the missing contributions, we replace $s$ by $u$ in Eqs.~(\ref{amw}) and (\ref{adsym}),
obtaining the
amplitudes ${A'}_{\gamma\gamma}^{(M)}$ and ${A'}_{\gamma\gamma}^{(D)}$. Adding them to $A_{\gamma\gamma}^{(M)}$
and $A_{\gamma\gamma}^{(D)}$ respectively, we arrive at
the complete expressions for the DLA amplitude of the process (\ref{gamma}) in the collinear kinematics.
In the Moderately-Virtual region (\ref{modoffshell}) it is
\begin{equation}\label{amcol}
\widetilde{A}_{\gamma\gamma}^{(M)} = A_{\gamma\gamma}^{(M)} + {A'} _{\gamma\gamma}^{(M)}
\end{equation}
whereas in the Deeply-Virtual region (\ref{deepy})

\begin{equation}\label{adcol}
\widetilde{A}_{\gamma\gamma}^{(D)} = A_{\gamma\gamma}^{(D)} + {A'} _{\gamma\gamma}^{(D)}.
\end{equation}

\section{Non-collinear photon-photon scattering}

In this Sect. we extend the results obtained above to the forward Regge kinematics (\ref{fkin}) with $t \neq 0$.
In order to avoid confusing new amplitudes with $A_{\gamma\gamma}^{(M)}$ and $A_{\gamma\gamma}^{(D)}$ obtained
under assumption that $t \sim 0$, we introduce a generic notation $M_{\gamma\gamma}$  for new amplitudes in DLA
and will provide this notation
with superscripts to specify the kinematics. The Born amplitude, $M_B$ is
(cf. Eq.~(\ref{aborn}))

\begin{equation}\label{mb}
M_B = \widetilde{B} + \widetilde{B}',
\end{equation}
with
\begin{equation}\label{mborn}
\widetilde{B} = - \imath \frac{e^4}{16 \pi^3}\int
\frac{d \alpha d \beta d k^2_{\perp}  w^2 k^2_{\perp}}
{k^2 (k + q - q')^2\left(x_q w + \beta w - \alpha x_q w + k^2\right) \left(x_p w + \alpha w - \beta x_p w + k^2 \right)}
\end{equation}
and $\widetilde{B}'$ is obtained from $\widetilde{B}'$ with replacing $q \to - q$.
It is obvious that the integration over $k$ in Eq.~(\ref{mborn}) yields a DL contribution from the region
\begin{equation}\label{kt}
k^2_{\perp} \gg -t = -(q'-q)^2.
\end{equation}

In other words, $|t|$ acts in Eq.~(\ref{mborn}) as a new IR cut-off. Therefore in the Born approximation (and beyond it)
the amplitude $M_{\gamma\gamma}$ in DLA is IR stable. All results we obtained in the previous Sects., studying
the amplitudes in the collinear kinematics, can easily be extended to the region of non-zero $t$ by the simple replacement

\begin{equation}\label{mut}
\mu^2 \to |t|.
\end{equation}

A further advancement strongly depends on the hierarchy between $Q^2_{1,2}$ and $|t|$. When

\begin{equation}\label{q12small}
Q^2_{1,2} < |t|,
\end{equation}
amplitude $M_B$ does not depend on $Q^2_{1,2}$ under the DL accuracy. In this case

\begin{equation}\label{qsmall}
M_B = M_B = -\frac{e^4}{16 \pi^2} \ln^2 \left(s/|t|\right).
\end{equation}

When $s \gg Q^2_{1,2} \gg |t|$, there are again two cases:

\begin{equation}\label{mbm}
M_B = -\frac{e^4}{16 \pi^2} \left[\ln^2 \left(s/|t|\right) -  \ln^2 \left(Q^2_1/|t|\right) - \ln^2 \left(Q^2_2/|t|\right)\right],
\end{equation}
when $Q^2_1 Q^2_2 \ll s |t|$
 and

\begin{equation}\label{mbd}
M_B = -\frac{e^4}{8 \pi^2} \ln^2 \left(s/Q^2_1\right) \ln \left(s/Q^2_2\right),
\end{equation}
when $Q^2_1 Q^2_2  \gg s |t|$. Now let us to extend our results for amplitudes $A_{\gamma \gamma}$ to
the case of non-zero $|t|$ beyond the Born
approximation. To this end, we
introduce new  logarithmic variables $\bar{\rho}, \bar{y}_1, \bar{y}_2$ instead of the variables
$\rho, y_1, y_2$ defined in Eq.~(\ref{y12}):

\begin{equation}\label{redef}
\bar{\rho} = \ln \left(s/|t|\right),~ \bar{y}_{1,2} = \ln \left(Q^2_{1,2}/|t|\right).
\end{equation}

In the case when $s \gg Q^2_1 \gg |t|$ and $Q^2_2 \lesssim |t|$
 i.e. when $\bar{\rho} > \bar{y}_1 > \bar{y}_2 \approx 0$, amplitude

\begin{equation}\label{mgammatildeqq}
\widetilde{M}_{\gamma\gamma}(\omega,y_1) =
\int_{-\imath \infty}^{\imath \infty} \frac{d \omega}{2 \pi \imath} e^{\omega (\bar{\rho} - \bar{y}_1)}
~\kappa
\left[f_0 (\omega) e^{y_1 f_0 /(8 \pi^2)} - \frac{a_0}{\omega} \right].
\end{equation}

In the more involved case of Moderate Virtualities, when $s \gg Q^2_1, Q^2_2 \gg |t|$ but $s |t| \gg Q^2_1 Q^2_2 $,
i.e. when
\begin{equation}\label{mbarkin}
 \bar{\rho}  >  \bar{y}_1 +  \bar{y}_2,
\end{equation}
the scattering amplitude $M_{\gamma\gamma}^{(M)}$ is

\begin{equation}\label{mmw}
M_{\gamma\gamma}^{(M)} =
\int_{- \imath \infty}^{\imath \infty} \frac{d \omega}{2 \pi \imath}
e^{\omega \bar{\rho}}~\kappa
 \left[ \bar{W}_1 + \bar{W}_2 \right],
\end{equation}
with

\begin{eqnarray}\label{wmbar}
\bar{W}_1 &=& e^{-\omega (\bar{\xi} +|\bar{\eta}|)/2}
\left(- \frac{a_0}{\omega} + f_0 e^{|\bar{\eta}| H}\right)
\\ \nonumber
\bar{W}_2 &=& \left(f_0 - \frac{a_0}{\omega}\right)
\frac{\omega}{\sqrt{\omega^2 - a_0/(2 \pi^2)}} \left[e^{-(\bar{\xi} + |\bar{\eta}|)\omega/2 + |\bar{\eta}| H} - e^{\bar{\xi} (-\omega + H)}\right].
\end{eqnarray}

whereas in the new Deeply-Virtual region

\begin{equation}\label{dbarkin}
 \bar{\rho}  <  \bar{y}_1 +  \bar{y}_2
\end{equation}
the scattering amplitude $M_{\gamma\gamma}^{(D)}$ is

\begin{eqnarray}\label{md}
M_{\gamma\gamma}^{(D)} &=&  \int_{- \imath \infty}^{\imath \infty} \frac{d \omega}{2 \pi \imath}
e^{\omega (\bar{\rho} - \bar{\xi}/2 - |\bar{\eta}|/2)}~\kappa
\left[
\left(- \frac{a_0}{\omega} + f_0 e^{|\bar{\eta}| H}\right)
+ \left(f_0 - \frac{a_0}{\omega}\right)
\frac{\omega}{\sqrt{\omega^2 - a_0/(2 \pi^2)}} e^{|\bar{\eta}| H}
\right].
\end{eqnarray}

Replacing $s$ by $u$ in Eqs.~(\ref{mmw}) and (\ref{md}), we obtain amplitudes ${M'}_{\gamma\gamma}^{(M)}$ and
${M'}_{\gamma\gamma}^{(D)}$. Adding them to the expressions in Eqs.~(\ref{mmw},\ref{md})
and multiplying them by the factor $2 n_f$, with $n_f$ being the number of involved flavors, we arrive at the
DL expressions for the scattering amplitudes $\widetilde{M}_{\gamma\gamma}^{(M)}$ (in the Moderately-Virtual region
(\ref{mbarkin}))
and $\widetilde{M}_{\gamma\gamma}^{(D)}$ (in the Deeply-Virtual region
(\ref{dbarkin})):

\begin{equation}\label{mmd}
\widetilde{M}_{\gamma\gamma}^{(M)} = 2 n_f [M_{\gamma\gamma}^{(M)} + {M'}_{\gamma\gamma}^{(M)}],~~
\widetilde{M}_{\gamma\gamma}^{(D)} = 2 n_f [M_{\gamma\gamma}^{(D)} + {M'}_{\gamma\gamma}^{(D)}].
\end{equation}

Amplitudes $\widetilde{M}_{\gamma\gamma}^{(M,D)}$ in Eq.~(\ref{mmd}) account for the total resuimmation
of DL corrections ot the Born amplitude $A_B$ of Eq.~(\ref{ab}) in the forward kinematic region (\ref{fkin}).

\section{Discussion of the obtained results}

In this Sect. we discuss the results obtained in the previous Sects.

\subsection{Comment on Deeply-Virtual and Moderately-Virtual regions}

The Deeply-Virtual (DV) and Moderately-Virtual (MV) kinematics are introduced in Eqs.~(\ref{smallq},\ref{modoffshell}) and
(\ref{bigq},\ref{deepy}) respectively. The scattering amplitude $A_{\gamma\gamma}^{(M)}$, being
calculated in MV kinematics explicitly depends on the IR cut-off $\mu$. Often, $\mu$
is an artificial parameter with arbitrary value but on the other hand,
there are cases, when
$\mu$ has a physical meaning. For instance, it can be the heavy quark mass or the masses of
$W,Z$ bosons in Standard Model. In such
cases
the range of $\rho$ in the region (\ref{deepy}) is quite restricted:

\begin{equation}\label{deepreg}
\max [y_1, y_2] < \rho < y_1 + y_2.
\end{equation}
As a result,
$A_{\gamma\gamma}^{(D)}$ cannot be used
for calculating the asymptotics of $A_{\gamma\gamma}$,
when $s \to \infty$. The asymptotics in this case can be obtained from $A_{\gamma\gamma}^{(M)}$. The same is true also
in the case when the running coupling effects for amplitudes in the Regge kinematics are accounted for.

On the contrary, when $\mu$ is not associated with an appropriate mass scale and $\alpha_s$ is fixed
or regarded as $\mu$- independent,
the value of $\mu$ can be chosen arbitrary small.
and, as the DV region ensures the IR stability independently of $\mu$, one can choose
$\mu$ very small. This considerably
broadens the applicability region for the DV kinematics and makes possible to use it at very high energies.
So, despite the kinematics is the Regge one, it is as IR stable as
the hard kinematics.

\subsection{Impact of Higher-loop DL radiative corrections}

The $s$-dependent parts, $B_{\gamma\gamma}^{(M,D)}$ of the photon-photon scattering amplitudes in the lowest-order approximation
at $t = 0$ are given by
Eqs.~(\ref{born1},\ref{born2}) for the MV and DV photons respectively. Accounting for the radiative corrections in DLA
converts them into amplitudes $A_{\gamma\gamma}^{(M,D)}$. Let us estimate the impact of the DL radiative
corrections on $B_{\gamma\gamma}^{(M,D)}$ in the simplest case when $Q^2_1 \sim Q^2_2 = Q^2$,
so the involved amplitudes in this case depend on $x = Q^2/s$ only. In
order to be independent of choice of $\mu$, we will do it for the amplitudes with Deeply-Virtual photons,
where $A_{\gamma\gamma}^{(D)} (x)$ is given by Eq.~(\ref{adq}).
We define the
the ratio $R_{\gamma}$ as follows:

\begin{equation}\label{rgamma}
R_{\gamma}(x)= A_{\gamma\gamma}^{(D)}/B_{\gamma\gamma}^{(D)}
\end{equation}
and show the plot of $R_{\gamma}$ against $x$ in Fig.~5.

\begin{figure}[h]
  \includegraphics[width=.4\textwidth]{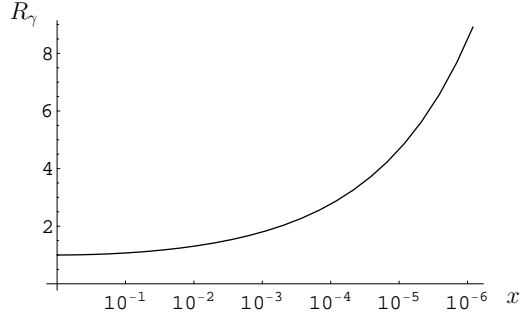}
  \caption{\label{gammafig5} Dependence of $R_{\gamma}$ on $x$.}
\end{figure}

Fig.~5 explicitly demonstrates that the radiative corrections become sizable since $x \sim 10^{-3}$.
They double the Bon amplitude at $x \sim 10^{-4} $.

\subsection{High-Energy asymptotics}

 At very high energies scattering amplitudes are often approximated by their asymptotics. The asymptotics
 are much easier to work on than the explicit expressions. However, the applicability region for the
 asymptotics cannot be deduced from theoretical consideration. Below we first show how to calculate
 the asymptotics and then outline  its applicability region.
  very small $x$
Asymptotics at $\rho \to \infty$ of all amplitudes we calculated above can be obtained
by using the saddle-point method and is given by similar expressions.
For the sake of simplicity we consider the small-$x$ asymptotics of amplitudes  $A_{\gamma\gamma}^{(M,D)}$
of Eqs.~(\ref{amw},\ref{adsym}) in the particular case when
$Q^2 \sim Q^2_2 \equiv Q^2$. In this case Eq.~(\ref{amw}) is reduced to

\begin{equation}\label{amq}
A_{\gamma\gamma}^{(M)} (x,y) =  \int_{- \imath \infty}^{\imath \infty} \frac{d \omega}{2 \pi \imath} x^{-\omega}
~\kappa
\left(f_0 - \frac{a_0}{\omega}\right)
\left[1 + \frac{\omega}{\sqrt{\omega^2 - a_0/(2 \pi^2)}}\left(1 - e^{-y \sqrt{\omega^2 - a_0/(2 \pi^2)}}\right)\right],
\end{equation}
where $x = Q^2/s$ and $y = Q^2/s$. In contrast, $A_{\gamma\gamma}^{(D)} $ depends on $x$ only:

\begin{equation}\label{adq}
A_{\gamma\gamma}^{(D)} (x) =
\int_{- \imath \infty}^{\imath \infty} \frac{d \omega}{2 \pi \imath} x^{-\omega}~\kappa
\left(f_0 - \frac{a_0}{\omega}\right) \left[1 + \frac{\omega}{\sqrt{\omega^2 - a_0/(2 \pi^2)}}\right].
\end{equation}

The saddle-point method, being applied to Eqs.~(\ref{amq},\ref{adq}), immediately yields that the rightmost stationary point $\omega_0$ is
the same for the both amplitudes and it is given by
the rightmost root of the equation
\begin{equation}\label{omega0}
\omega^2 - a_0/(2 \pi^2) = 0.
\end{equation}

The further progress depends on the treatment of $\alpha_s$.
 When $\alpha_s$ is fixed ($\alpha_s = \alpha_s^{fix}$), it is easy to obtain an analytic expressions for $\omega_0$:
 \begin{equation}\label{omega0fix}
 \omega_0 = \sqrt{2\alpha_s^{fix} C_F/\pi} + 1/(2 z).
 \end{equation}
A numerical estimate for $\alpha_s^{fix}$ in Eq.~(\ref{omega0fix}) was obtained
in Ref.~\cite{egtfrozen}. According to it,  $\alpha_s \approx 0.24$.

When the running coupling effects are taken into account, $a_0$ depends on $\omega$ (see Eq.~(\ref{aqqrun}))
and therefore Eq.~(\ref{omega0}) has to be solved numerically (see Ref.~\cite{egtquark}). It leads to the estimate
$\omega_0 \approx 0.4$.
The small-$x$ asymptotics $\left[A_{\gamma\gamma}^{(M,D)}\right]_{as}$ of amplitudes $A_{\gamma\gamma}^{(M,D)}$
respectively are:

\begin{equation}\label{amqas}
\left[A_{\gamma\gamma}^{(M,D)}\right]_{as} = \Pi^{(M,D)} x^{- \omega_0},
\end{equation}
with the factors $\Pi^{(M)}$ and $\Pi^{(D)}$ being

\begin{equation}\label{pimd}
\Pi^{(M)} = \frac{e^4}{\pi^2 \omega_0^3} \frac{1}{\sqrt{2 \pi \omega_0 z^3}}
\left[1 + \frac{\omega_0 z}{2} \left(1 - e^{-2y/z}\right)\right],~~
\Pi^{(D)} = \frac{e^4}{4 \pi^2 \omega_0^2} \frac{1}{\sqrt{\pi \omega_0 z}},
\end{equation}
where we have denoted $z = \ln (1/x)$.

Now let us consider the $x$ dependence of the ratio

\begin{equation}\label{rasm}
  R_{as} = \frac{\left[A_{\gamma\gamma}^{(M)}\right]_{as}}{ A_{\gamma\gamma}^{(M)}}.
\end{equation}

The $x$-dependence of $R^{(M)}_{as}$  is plotted in Fig.~5. For the sake of simplicity, the graph in Fig.~6 is
done for $y \approx 0$. The greater $y$, the lower the graph runs.

\begin{figure}[h]
  \includegraphics[width=.4\textwidth]{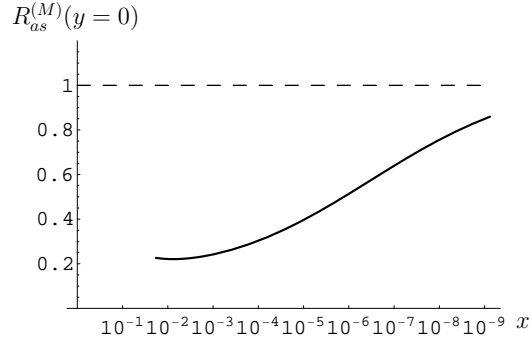}
  \caption{\label{gammafig6} Dependence of $R^{(M)}_{\gamma}$ on $x$.}
\end{figure}

Similarly to Eq.~(\ref{rasm}), we define the ratio

\begin{equation}\label{rasd}
  R_{as} =
  \frac{\left[A_{\gamma\gamma}^{(D)}\right]_{as}}{ A_{\gamma\gamma}^{(D)}}.
\end{equation}
The $x$-dependence of $R^{(D)}_{as}$ is plotted in Fig.~7.

\begin{figure}[h]
  \includegraphics[width=.4\textwidth]{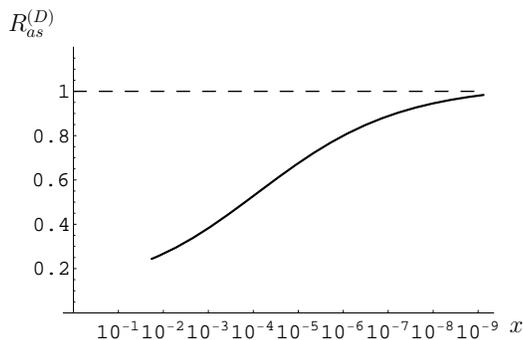}
  \caption{\label{gammafig7} Dependence of $R^{(D)}_{\gamma}$ on $x$.}
\end{figure}

Figs.~6,7 display that the amplitudes $A_{\gamma\gamma}^{(M,D)}$ are reliably represented by their asymptotics at very small $x$.
Indeed, $R_{as}^{(M,D)} \sim 0.8$ at $x \lesssim 10^{-8}$. It perfectly agrees with the results of Ref.~\cite{egtsum} where it
was proved that the small-$x$ asymptotics of the non-singlet DIS structure functions $F_1^{NS}$ and $g_1^{NS}$
reliably represent these structure functions at $x \lesssim 10^{-8}$. In terms of the Reggeology, the only
difference between $F_1^{NS}$ and $A_{\gamma\gamma}^{(M,D)}$ is the difference in the impact-factors
while the Reggeons are the same. This proves that
the applicability region for the use of non-vacuum Reggeons is

\begin{equation}\label{regx}
x \lesssim 10^{-8},
\end{equation}
i.e. Eq.~(\ref{regx}) manifests that the non-vacuum Reggeons should not be used for description of hadronic reactions at available energies.

\section{Conclusion}

We have calculated the amplitudes of the process (\ref{gamma}) in DLA. We considered this process in both the collinear
kinematics (amplitudes $A_{\gamma\gamma}^{(M)}$ and $A_{\gamma\gamma}^{(D)})$, where $t=0$, and  the forward kinematics (\ref{fkin})
(amplitudes $M_{\gamma\gamma}^{(M)}$ and  $M_{\gamma\gamma}^{(D)} $)
at  $t \neq 0$. So as to calculate those amplitudes we constructed
and solved the Infrared Evolution Equations for them.  According to the general technology of solving IREE, any
general solution to IREE for a scattering amplitude in a certain kinematics
is specified through matching with the known amplitude of the same process in a simpler kinematics.
So, before solving IEEE for the amplitudes  $M_{\gamma\gamma}^{(M,D)}$ in the $t \neq 0$ -kinematics, we
had to calculate the amplitudes $A_{\gamma\gamma}^{(M,D)}$ of the same process in the collinear kinematics. Doing so, we
confirmed the results of Ref.~\cite{bl} and generalized them to the case of running  QCD coupling
while $\alpha_s$ in Ref.~\cite{bl} was fixed.

At very high energies the scattering amplitudes are often considered in the asymptotic form and
such asymptotics are addressed as Reggeons. Such Reggeons are much easier to use than their parent
amplitudes. However, applicability regions for the asymptotics (Reggeons) cannot be fixed from
theoretical grounds. We do it numerically, calculating the asymptotics of
amplitudes $A_{\gamma\gamma}^{(M,D)}$ and comparing the asymptotics to the amplitudes.
In order to calculate the asymptotics,
we use the saddle-point method. The results are plotted in Figs.~5,6. They outline the applicability
region of the non-vacuum Reggeons and perfectly agree with the
observation of Ref.~\cite{egtsum} that non-vacuum Reggeons should not be used for describing
available experimental data.

\section{Acknowledgements}

We are grateful to W.~Schafer and A.~Szczurek for useful communications. 
The work of D.Yu. Ivanov was supported by the program of fundamental scientific
          researches of the SB RAS № II.15.1., project № 0314-2016-0021

\end{document}